\documentclass[prd,aps,
superscriptaddress,preprintnumbers,nofootinbib,showpacs]{revtex4}

\usepackage{graphicx,wrapfig}
\usepackage{subfigure}
\usepackage{amssymb}
\usepackage{amscd}
\usepackage{amsfonts}
\usepackage{amsmath}
\usepackage{exscale}
\usepackage{pslatex}
\usepackage{graphicx,color}

\newcommand{\vx}{\vec{x}}
\newcommand{\vy}{\vec{y}}
\newcommand{\vk}{\vec{k}}
\newcommand{\vp}{\vec{p}}
\newcommand{\mM}{\mathcal{M}}
\newcommand{\mN}{\mathcal{N}}
\newcommand{\mV}{\mathcal{V}}

\newcommand{\mG}{\mathcal{G}}
\newcommand{\p}{\partial}

\newcommand{\beqa}{\begin{eqnarray}}
\newcommand{\eeqa}{\end{eqnarray}}
\newcommand{\beq}{\begin{eqnarray}}
\newcommand{\eeq}{\end{eqnarray}}

\begin{document}

\title{Origin of Confining Force}

\author{Patrick~Cooper}
\email[]{pjc370@nyu.edu}
\affiliation{New York University, New York, NY 10003}

\author{Daniel~Zwanziger}
\email[]{dz2@nyu.edu}
\affiliation{New York University, New York, NY 10003}

\date{Dec. 17, 2015}

\begin{abstract}

        In this article we present exact calculations that substantiate a clear
        picture relating the confining force of QCD to the zero-modes of the
        Faddeev-Popov (FP) operator $\mM(gA) = - \p \cdot D(gA)$.  This is done
        in two steps.  First we calculate the spectral decomposition of the FP
        operator and show that the ghost propagator $\mG(k; gA) = \langle \vk |
        \mM^{-1}(gA) | \vk \rangle$ in an external gauge potential $A$ is
        enhanced at low $k$ in Fourier space for configurations $A$ on the
        Gribov horizon. This results from the new formula in the low-$k$ regime
        $\mG^{ab}(k,gA) = \delta^{ab} \lambda_{|\vk|}^{-1}(gA)$, where
        $\lambda_{|\vk|}(gA)$ is the eigenvalue of the FP operator that emerges
        from $\lambda_{|\vk|}(0) = \vk^2$ at $A$ = 0. Next we derive a strict
        inequality signaling the divergence of the color-Coulomb potential at
        low momentum $k$ namely, $\widetilde{\mV}(k) \geq k^2 \mG^2(k)$ for $k
        \to 0$, where $\widetilde{\mV}(k)$ is the Fourier transform of the
        color-Coulomb potential $\mV(r)$ and $\mG(k)$ is the ghost propagator in
        momentum space. Although the color-Coulomb potential is a gauge-dependent
        quantity, we recall that it is bounded below by the
        gauge-invariant Wilson potential, and thus its long range provides a
        necessary condition for confinement.  The first result holds in the
        Landau and Coulomb gauges, whereas the second holds in the Coulomb
        gauge only. 
\end{abstract}

\maketitle

\section{Introduction}%
     
Experimentally, color confinement is the most well tested phenomenon of QCD -
verified every time a free quark from a high energy collision doesn't make it
to the calorimeters without bounding itself up into hadrons. Nevertheless,
theoretically we lack a satisfying mechanism to reveal this feature. This is
not to say our theory of strong interactions is incapable of describing
confinement. When the QCD partition function is simulated on a spacetime
lattice, gauge-invariant observables can be computed, revealing a confining
gluonic flux tube between colored states. These simulations reveal a picture of
a string with a tension (energy per length) that to leading order yields a
linearly rising potential asymptotically at long distances. An \emph{analytic}
method that could show the origin of this confining force has long been sought
after by generations of field theorists.

Understanding the origin of forces between particles in QCD is significantly more
challenging than its abelian cousin because quantities that were once gauge
\emph{invariant} like the curvature of the gauge connection, are now gauge
\emph{covariant}.  Thus, even the color-electric field is a gauge-dependent
quantity so the naive notion of ``force" that one might have in QCD which would
follow from a non-abelian Lorentz force law is more subtle. This is a situation
much like gravity, where observables that were once gauge invariant (local
scalar fields) become gauge covariant with respect to diffeomorphisms. The root
of this complication is that in non-abelian theories, the kinetic terms
introduce non-linear interactions between the force carriers themselves, and
thus the color-electric field itself carries charge. 

The first crucial result that we make use of in this article, allowing one to
extract gauge-invariant information from a gauge dependent quantity, is the
theorem ``no confinement without Coulomb confinement" from
\cite{Zwanziger:2002sh}. This theorem tells us that a necessary condition for
confinement is that the color-Coulomb potential $\mV(r)$ be confining via the
inequality
\beq
V_W(r) < - C \mV(r) \; \; \; r \to \infty,
\eeq
where $V_W(r)$ is the Wilson potential, ie the contribution to the energy coming
from the flux tube formed between two quarks that furnish some representation
with Casimir $C$.

$\mV(\vx)$ is defined to be the instantaneous part of the time-time component
of the gluon propagator in Coulomb gauge \cite{Cucchieri:2000hv}
\beq
    \label{CCP}
    \langle A_0^a(x) A_0^b(y) \rangle = \delta^{ab} [ \mV(\vx-\vy)
    \delta(x_0 - y_0) - P(x-y) ],
\eeq
where $P(x-y)$ is non-instantaneous.  This formula is obtained via the
first-order formalism in Coulomb gauge \cite{Baulieu:1998kx}. After integrating
out $A_0$ and the longitudinal component of the conjugate momentum, $\pi_i$,
and resolving Gausses law, the partition function becomes
\beq
        Z(J_0) = \int \mathcal D \pi^{\mathrm{tr}}_i \mathcal D A^{\mathrm{tr}}_i
        \mathrm{exp} \left[ \int d^Dx \left(  i
\pi^{\mathrm{tr}}_i \dot A_i^{\mathrm{tr}} -\frac{1}{2} (\pi^{\mathrm{tr}}_i)^2 -
                \frac{1}{2} B_i^2 - \frac{1}{2} \rho \left( {\mM}^{-1}(-\p)^2 {\mM}^{-1} \rho \right) \right) \right]
\eeq
with the color-charge density $\rho$ given by
\beq
        \rho = \rho_0 + \rho_{dyn} = J_0^a + g f^{abc} A_i^{b \, \mathrm{tr}}
        \pi_i^{c \, \mathrm{tr}},
\eeq
the superscript $\mathrm{tr}$ indicates a transverse 3-vector, and $B_i$ is the color-magnetic field.  Here $ ({\mM}^{-1} )^{ab}(\vx, \vy; gA)$ is the ghost propagator
in Coulomb gauge in a transverse background potential $A_i^b$, with $\p_i
A_i^b = 0$,
\beq
\mM^{ac} \equiv - D^{ac} \cdot \p = \mM_0^{ac} + \mM_1^{ac}
= (- \p^2)\delta^{ac}  - g f^{abc} A_i^b \p_i
\eeq
is the
Faddeev-Popov operator in a space of dimension $d$, where $A_i$ is transverse
$\p_i A_i = 0$, so $\mM$ is hermitian, and $i = 1, ... d$.  The
indices $a, b, c = 1, ...  N^2-1$ take values in the adjoint representation of
the $SU(N)$ color group.  In Coulomb gauge $d = 3$ is the dimension of the
ordinary 3-dimensional space part of 4-dimensional Euclidean or Minkowskian
space. Those of our results which concern the ghost propagator, hold also in
Landau gauge, in which case $d$ is the dimension of Euclidean space-time.  By taking variational derivatives of $Z(J_0)$ and setting $J_0=0$, we get 
\beq
\label{v_eq}
  \delta^{ab}  \mV(\vx-\vy) =  \left< (\mN^{-1})_{\vx \vy}^{ab} \right> 
\eeq
and 
\beq
        \delta^{ab} P(x-y) = \left< ( \mN^{-1}\rho_{dyn})_x^a  (\mN^{-1}\rho_{dyn})_y^b \right>
\eeq
where 
\beq
\label{defineNinverse}
\mN^{-1} = \mM^{-1} (-\p^2) \ \mM^{-1}.
\eeq
The quantity $ ( {\mN}^{-1} )^{ab}(\vx, \vy; gA)$ is the color-Coulomb potential in a
transverse background potential $A_i^b$. To all orders in perturbation theory
$g^2 \mV(\vx-\vy)$ is a renormalization-group invariant
\cite{Zwanziger:1998ez}. That is, the color-Coulomb interaction energy, $g^2
\mV(\vx-\vy)$, is scheme-independent, and if this is also true
non-perturbatively than $g^2 \mV(\vx-\vy)$ has a well defined value in Joules
at any given separation $|\vx - \vy|$ in meters.

An important point is that both the instantaneous part $\mV(\vx - \vy)$ and the non-instantaneous part $P(x-y)$ are each the kernel of a positive operator, so the opposite signs in \eqref{CCP} mean that $P(x-y)$ is screening while $\mV(\vx - \vy)$ is anti-screening.  Which one dominates is a dynamical question that we will not address here.  However while a \emph{sufficient} condition for confinement is the ideal, one can at least obtain
a necessary condition by calculating the infrared behavior of the two-point instantaneous
correlator $\mV(\vx)$ which is the anti-screening part.

In this article we show by direct calculation that
$\mV(\vx-\vy)$ is in fact confining, granted a few not unreasonable assumptions
about certain generic features of the spectrum of the Faddeev Popov (FP)
operator.  An important tool
in this analysis is the implementation of the non-local `horizon condition'
\cite{Zwanziger1989513} which restricts the gluon measure to the first Gribov
region \cite{GRIBOV19781}. It should come as no surprise that the Gribov
ambiguity (which is the statement that no continuous, finite-energy, globally
defined gauge connection exists over spacetime) has a role to play in the long
distance behavior of YM theory. In \cite{Singer:1978dk} Singer showed that this
problem is a logical consequence of the highly non-trivial topology of the
space of connections over spacetime quotiented by the group action of any
non-abelian group. Since this obstruction is topological, perturbation theory
in a neighborhood of a flat connection is unaffected and thus the ambiguity is
relegated to a merely academic discussion in most textbooks on field theory
which mostly seek perturbative treatments of QCD. To probe very large length
scales of the gluon field however, we must concern ourselves with a
\emph{global} section of the SU(3) principle fibre bundle over spacetime, and
thus the non-trivial topology of the bundle can be felt. Indeed it was shown
immediately in Gribov's founding paper \cite{GRIBOV19781} that this in fact
happens. In ameliorating this ambiguity, the low $k$ modes of the gluon
propagator go from IR singular to IR suppressed, and the pole is removed from
the real axis, thus removing the gluon from the physical spectrum. With the
gluon suppressed, in the Gribov confinement scenario, it's the enhancement of
the ghost propagator that is intuited as responsible for the long-range force. 

Other scenarios also relate this treatment of the Gribov ambiguity to
confinement. A recent, very elegant picture was provided by Reinhardt in
\cite{Reinhardt:2008ek} to show that the `no-pole' condition, an equivalent
statement to the aforementioned `horizon condtion', implies that the QCD
vacuum is a perfect color dielectric ($\epsilon(k) = 0$)  resulting in a dual
Meissner effect, confining electric flux into vortices, which due to dynamical
quark pair creation can never be macroscopic in size.

Our approach will be to realize the concrete connection between the horizon
condition and the divergence of the color-Coulomb potential (CCP). This is done by
using scattering theory techniques to locate the Gribov horizon. We will also
use these results to prove a specific asymptotic relationship between 
a lower bound for the CCP and the ghost propagator, namely
\beq
\label{inequality}
\widetilde{\mV}(k) \geq k^2 \mG^2(k) \ \ \ \ \ \  {\rm for} \  k \to 0,
\eeq
 where $\widetilde{\mV}(k)$ is the Fourier transform of the color-Coulomb
 potential $\mV(r)$, and $\mG(k)$ is the ghost propagator in momentum space.
 This makes explicit the role of the enhanced ghost propagator in
 confinement.\footnote{Subtleties regarding level crossing and discreteness of
         the FP spectrum in the infinite-volume limit could in principle poke
 holes in some of the conclusions, however generically they would be
 unexpected.}  Since $\mV(\vx-\vy)$ is gauge-dependent, this result is not a
 proof of confinement and, in fact, the non-instantaneous force $P(x)$ in
 \eqref{CCP} may compensate the instantaneous force, as happens in the
 deconfinement transition of the gluon plasma at sufficiently high temperature,
 or when dynamical quarks screen external quarks. However it reverses the
 question of confinement. The problem now is not to show that a confining force
 exists; that is always provided by the instantaneous color-Coulomb force. The
 problem is to determine whether or not this force is compensated by
 non-instantaneous forces. This provides a very intuitive theory of
 confinement, quite analogous to our theory of atoms or molecules which are
 themselves neutral, but are held together by instantaneous Coulomb forces
 between electrically charged constituents.  In this article we have ignored the presence of dynamical quarks because we do not expect that they will screen the color-Coulomb confining force.  Indeed we expect that although dynamical quarks do screen the gauge-invariant Wilson potential nevertheless, in the picture provided by the Coulomb gauge, the color-Coulomb potential always acts between elementary color-charges --- be they quarks or gluons --- just as the electro-static Coulomb potential always acts between elementary electric charges.  Indeed it is the presence of the color-Coulomb potential that makes it energetically favorable for color-neutral particles to be formed.
 
In the final sections, we'll discuss what can be said about the gauge-invariant
aspects of this analysis. We'll also see how our simple picture holds up next
to recent lattice studies of the relevant parts of the FP spectrum. 

\section{The Spectral Decomposition of the FP operator: Results}%

Gribov showed that covariant gauges weren't sufficient to fix the gauge
redundancy of the measure of the gluon field, and proposed a simple but
drastic reduction of the gluon measure of integration. The geometric picture is that
the gauge is simply the condition that the gluon field be a critical
point, $\delta F = 0$, of the functional given by the $L^2$ norm, 
\beq
    \label{gauge_functional}
    F(A) = \int d^dx|A(x)|^2,
\eeq
where the variation is done with respect to an infinitesimal gauge
transformation. With this condition alone, every local
minimum, maximum and saddle point of the same gauge orbit would be redundantly
counted in the path integral. Gribov proposed that the first thing one
can do to ameliorate this situation is demand that $A(x)$ be such that the FP
operator be positive definite. It can be shown that this is equivalent to
demanding that $A(x)$ be a local minimum of \eqref{gauge_functional} \cite{Maskawa:1978rf, Zwanziger:1981nu,Shanskii:1986}. Since the
inverse of the FP operator is the ghost propagator, Gribov did a semiclassical
computation to see where the ghost propagator went through zero away from the
usual pole at $k=0$ to derive a condition where the path integration should be
cut off in configuration space.  This cut-off has been shown to restrict the region of integration to a bounded convex region \cite{Zwanziger:1982na, Shanskii:1986}, and the way it
alters perturbation theory already starts to show qualitative features of
confinement. This approach to restricting the measure has been followed up by
various other methods \cite{Vandersickel:2012tz} that have lead to a local,
renormalizable Lagrangian with spontaneously broken BRST invariance including
new auxillary ghosts to implement this cutoff \cite{Zwanziger1989513}.  

In Appendix \ref{evalue_app}, we carefully revisit the heuristic approach
to deriving the non-local `horizon condition' which implements the cut off to
the Gribov region, found in \cite{Zwanziger1989513} using a method akin to
degenerate perturbation theory, but does not rely on any perturbative expansion. The idea is that the FP operator can be
considered as a deformation of the Laplacian, which is already positive
definite.  One slowly turns on $gA$ until one of the eigenvalues becomes negative. It
should be noted that this will be the way we generate an implicit equation for
the eigenvalues, and that at no point do we require $gA$ to be small as in
typical perturbation theory. We do however limit our results to
asymptotically small $k$ since it's the infrared region in which we're
interested. We designate by $\lambda_{|\vec k|}(gA)$ the eigenvalue of the FP
operator coming out of the eigenvalue of $\mathcal M_0$ at $k^2$ as we
adiabatically turn on the perturbation $\mathcal M_1(gA)$.

The details are carried out in Appendices \ref{evalue_app}, B, and C.  Here we'll
just present an abridged summary of the results so the reader can continue on
to the physics. The eigenvalue of the FP operator for a given connection near
$k=0$ in Fourier space is given by
\beq
    \label{evalue_eqn}
    \lambda_{|\vk|}(gA) = \vk^2 \left( 1 - [(N^2 - 1) d V]^{-1} H(gA) +
    j_{|\vk|}(gA) \right),
\eeq
where $j_{|\vk|}(gA)$ vanishes as $k$ tends to $0$. $H(gA)$ is the horizon
function given by 
\beq
    \label{horizonfunctioneq}
    H(gA) = \int d^dx \, d^dy D_{i \vx}^{a c} D_{i \vy}^{a e} \left( M^{-1}
    \right)^{c e}_{\vx, \vy} .
\eeq
The horizon condition \cite{Vandersickel:2012tz} and no-pole condition \cite{GRIBOV19781} both yield
\beq
    \label{horizon_condition}
    \left< H(gA) \right> = (N^2-1) d V.
\eeq
This condition enhances the ghost propagator $\mG(k)$ at momentum $k = 0$.  We define a new IR exponent $\eta$ as an ansatz for the behavior of the
ensemble average, $\left< \lambda^{-1}_k \right>$, which goes to zero like
$1/k^{2+\eta}$. As we'll see, much of the qualitative features of confinement
boil down to the value of this exponent. It can be calculated numerically on the lattice and
attempts at calculating it analytically make up a large part of the focus of recent studies
of long-range QCD using Schwinger Dyson Equations (SDE) and variational methods
\footnote{In the literature these recent papers are known as the ``T\"{u}bingen
approach", coined in \cite{Reinhardt:2011af}. See also \cite{Campagnari:2010wc,
Schleifenbaum:2006bq}}.

Another important result, is that while to order $k^2$ the eigenvalues are
strongly perturbed, the projector onto eigenstates of the FP operator
adiabatically emerging from the degenerate level $|\vec{k}|$ of the Laplacian
remains equal to the unperturbed projector.  Namely $P_{|\vec{k}|}(gA) =
P_{|\vec{k}|}(0) + O(\vec{k}^2)$ or in terms of the eigenstates of the FP
operator
\beq
        \lim_{\vk \to 0} \sum_{ \ \vk; |\vk |} | \psi_{\vk}(gA)
        \rangle \langle \psi_{\vk}(gA) | = \sum_{ \ \vk; |\vk |} |
        \vk \rangle \langle \vk |.
\eeq
(Here and below $\sum_{ \ \vk; |\vk |}$ means sum on all vectors $\vk = (2 \pi
\vec n/ L)$, with the same fixed $| \vk |$, where the $n_i$ are integers.)
By relating the enhanced ghost propagator $\mG(k)$ and the CCP $\widetilde \mV(k)$ to eigenvalues and projectors
onto the space of eigenvectors of the FP operator, we obtain the lower bound \eqref{inequality} on the IR strength of the CCP.

\section{The Long Range Color Coloumb Potential}%

In Coulomb gauge the propagator of the Faddeev-Popov ghost is instantaneous,
\beq
    \langle c^a(x) \bar c^b(y) \rangle = \delta^{ab} \mG(\vx - \vy)
    \delta(x_0 - y_0), 
\eeq
and is expressed in terms of the Faddeev-Popov operator by
\beq
    \delta^{ab} \mG(\vx-\vy) =  \langle ( {\mM}^{-1} )^{ab}(\vx,
    \vy; gA) \rangle.
\eeq

The two operators $\mM$ and $\mN$ are related by \cite{Feuchter:2004mk} 
\beq
\label{relateoperators}
    \mN^{-1} = (g \p_g + 1) {\mM}^{-1},
\eeq
where $\p_g \equiv \p / \p g$.  This follows from the elementary identity
\beq
    g \p_g {\mM}^{-1} = -  {\mM}^{-1}  {\mM}_1
    {\mM}^{-1} =  {\mM}^{-1}  ({\mM}_0 - \mM)
    {\mM}^{-1} = \mN^{-1} - \mM^{-1}.
\eeq
The spectral decomposition of the ghost propagator is given by
\beq
\label{expandghostprop}
    \left( \mM^{-1} \right)_{xy}^{ab} = \sum_n { \psi_{nx}^a
    \psi_{ny}^b \over \lambda_n },
\eeq
where we have used the fact that the Faddeev-Popov operator $\mM$ is a
real symmetric operator, and we have chosen eigenfunctions, $\mM \psi_n
= \lambda_n \psi_n$, that are real $\psi_{nx}^a = (\psi_{nx}^a)^*$.  From
\eqref{relateoperators} we obtain
\beq
\label{expandNinversea}
    \left( \mN^{-1} \right)_{xy}^{ab} = \sum_n \left[ \left( { - g
    \p_g \lambda_n \over \lambda_n^2 } + { 1 \over\lambda_n } \right)
    \psi_{nx}^a \psi_{ny}^b + {1 \over \lambda_n} g \p_g( \psi_{nx}^a
    \psi_{ny}^b) \right].
\eeq
We expand $g \p_g \psi_{nx}^a = \sum_m c_{nm} \psi_{mx}^a$, and observe that
\beq
    g \p_g \int d^dx \ \psi_{nx}^a \psi_{nx}^a = 2 \int d^dx \sum_m
    c_{nm} \psi_{mx}^a \psi_{nx}^a = 2 c_{nn} = 0 \hspace{2cm} {\rm ( no \ sum
    \ on \ } n),
\eeq
which vanishes because the eigenfunctions are normalized, $\int d^dx
\psi_{nx}^a \psi_{nx}^a = 1$. Thus the color-Coulomb potential
$\mN^{-1}$ in the background field $A_i^a$ has a kind of spectral
decomposition in terms of the eigenfunctions of the Faddeev-Popov operator,
\beq
\label{modeexpV}
    \left( \mN^{-1} \right)_{xy}^{ab} = \sum_n \left( { - g \p_g
    \lambda_n \over \lambda_n^2 } + { 1 \over\lambda_n } \right)  \psi_{nx}^a
    \psi_{ny}^b +  \sum_{n \neq m} \left( { c_{nm} \over \lambda_n} + { c_{mn}
    \over \lambda_m} \right) \psi_{mx}^a \psi_{ny}^b .
\eeq
Its trace,
\beq
\label{modeexpVa}
    \int d^dx \left( \mN^{-1} \right)_{xx}^{aa} = \sum_n \left( { - g
    \p_g \lambda_n \over \lambda_n^2 } + { 1 \over\lambda_n } \right),
\eeq
is expressed in terms of the eigenvalues $\lambda_n$ of the Faddeev-Popov
operator $\mM$.

\section{Necessary condition for confining potential}%

The color-Coulomb potential $\mV(\vx - \vy)$ is said to be confining if the
infrared contribution to the color-Coulomb self-energy $\mV(0)$ of any colored
particle is divergent \cite{Greensite:2004ke}. This is a necessary condition for
physical confinement because of the theorem ``no confinement without Coulomb
confinement," \cite{Zwanziger:2002sh} which provides a lower bound on the
color-Coulomb potential at large separation.  However it is not a sufficient
condition.

The color-Coulomb self-energy is given by
\beqa
\label{self-energy}
    \mV(0) & = & (N^2 -1)^{-1} \left\langle  (\mN^{-1})^{aa}(\vx, \vx; gA)
    \right\rangle 
    \nonumber \\
    & = & [(N^2 -1)V]^{-1} \left\langle \int d^dx \ (\mN^{-1})^{aa}(\vx, \vx; gA)
    \right\rangle.
\eeqa
Since this quantity is a trace it can be expressed in any basis.  Thus, from
\eqref{modeexpV}, we obtain
\beq 
    \mV(0) = { 1 \over (N^2 -1)V  } \sum_{|\vk| < k_{\rm max}}
    \left\langle  { - g \p_g \lambda_{|\vk|}(gA) \over \lambda_{|\vk|}^2(gA) }
    + {  1 \over \lambda_{|\vk|}(gA) } \right\rangle .
\eeq
 We have introduced a cut-off $k_{\mathrm{max}}$ to avoid the ultraviolet
divergence of the self-energy.  It is the infrared divergence of the
self-energy that is a necessary condition for a confining potential. Here
$\mV(0)$ is expressed entirely in terms of the eigenvalues
$\lambda_{|\vk|}(gA)$ of the Faddeev-Popov operator. Whether or not $\mV(0)$
is confining is determined by the eigenvalues and the density of eigenvalues in
the limit $\vk \to 0$, that is to say by the zero modes.

\section{Zero modes of Faddeev-Popov operator and confinement}%

\begin{figure}
    \begin{center}
    \includegraphics[width = 12 cm]{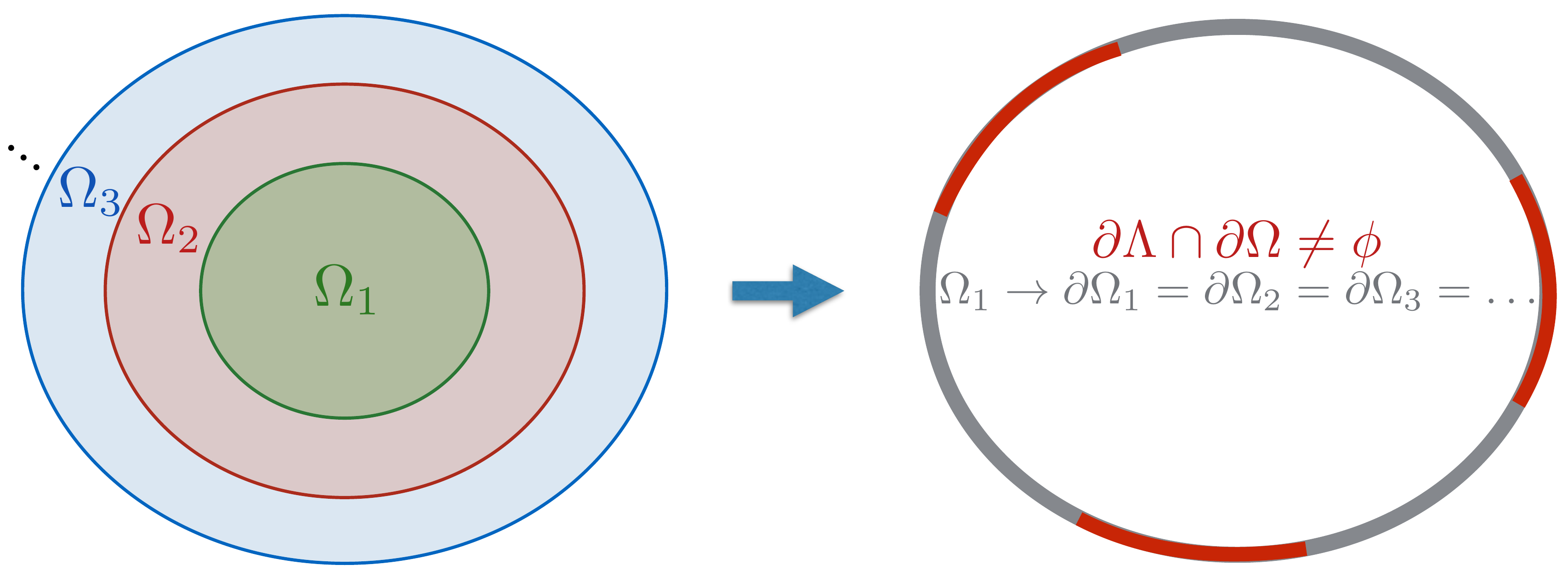}
    \caption{In light of this result, to lowest order in $k$, all levels cross
    the horizon at once, thus altering the traditional picture of the various
    Gribov regions to one that makes apparent this infinite density of
    states. $\Lambda$ is the fundamental modular region which corresponds to the
    absolute minimum of the functional \eqref{gauge_functional}.
    \label{one_horizon}}
    \end{center}
\end{figure}

\begin{figure}
    \begin{center}
    \includegraphics[width = 10 cm]{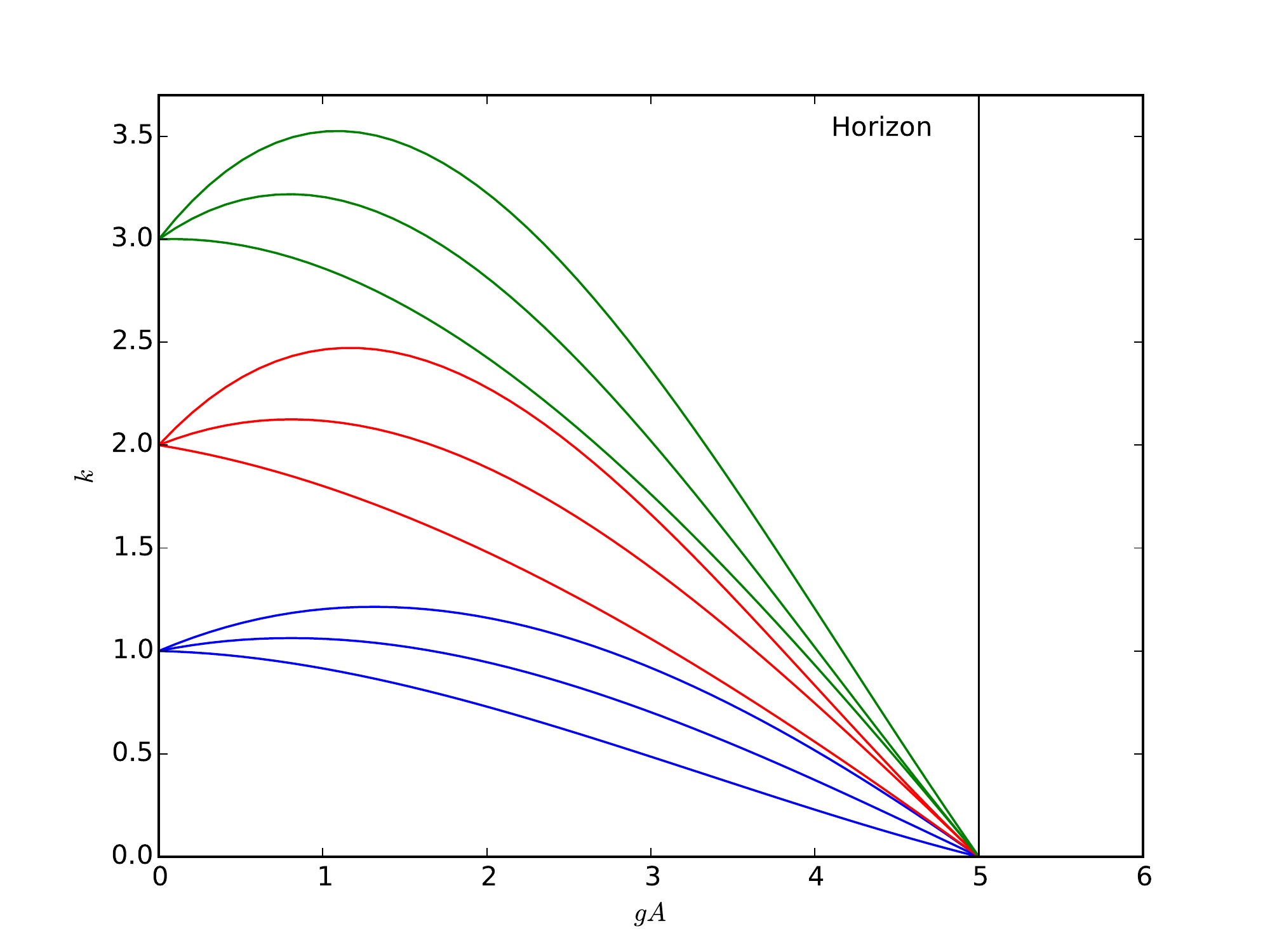}
    \caption{Above is a cartoon (a 1D projection of an infinite dimensional
            space) of the cubic symmetry degeneracy of the eigenvalues of the
            finite volume Laplacian being lifted by a perturbation and heading
            towards zero. Wave vectors are measured in units of $2 \pi/ L$.
            The vertical line marks the location of the horizon where they
            cross the trivial eigenvalue at 0.
    \label{perturbed_evalue}}
    \end{center}
\end{figure}

The near-zero eigenvalues of the Faddeev-Popov operator are studied in
Appendices \ref{evalue_app} and \ref{app_projector}. According to \eqref{evalue_eqn}, the eigenvalues are
given to order $\vk^2$ by
\beq
\label{factorksquarex}
    \lambda_{| \vk |}(gA) = \vk^2 \left( 1 - [d (N^2-1)V]^{-1} H(gA)
    \right).
\eeq
The crucial observation is that, to leading order in $\vk$, all eigenvalues
$\lambda_{| \vk |}(gA)$, for different $\vk = 2 \pi \vec{n} /L$, where the
$n_i$ are integers, pass through zero together in a massive level crossing, at
a common Gribov horizon as the result of \eqref{horizon_condition}. This is
illustrated in Figs.~\ref{one_horizon} and \ref{perturbed_evalue}. This
massive level crossing also occurs in lattice gauge theory on large lattices
as was noted in \cite{Zwanziger:1991ac} under the heading ``all horizons are one
horizon".

There are now two effects that together strongly
enhance the infrared self-energy. (i) The massive level crossing makes the
density of states, that is, the density of zero-modes, of the operator $\mM$
diverge as compared to the density of zero modes of the Laplacian operator.
This fits nicely with confinement scenario described in \cite{Greensite:2004bv,
Greensite:2004ur,Nakagawa:2010eh} where the divergence of the CCP is also understood as
resulting from an enhanced density of states. (ii) The ratio $\lim_{\vk \to 0}
\vk^2 / \lambda_{|\vk|} \to \infty$ diverges, which provides an enhancement of
the spectral decomposition of $\mN^{-1}$ over and above that of $\mM^{-1}$. 

We now establish some simple bounds that show intuitively how a long-range
color-Coulomb potential arises.  To leading order in $\vk$, the infrared
self-energy \eqref{self-energy} is given by
\beq
    \mV(0) \approx  { 1 \over (N^2 -1)d V^2 }  \sum_{|\vk| < k_{\rm max}}
    \left\langle  { g \p_g H(gA) \  \vk^2  \over \lambda_{|\vk|}^2(gA) }
    \right\rangle,
\eeq
where the $\approx$ symbol means `leading infrared behavior'.  According to \eqref{ehorizonfunction} we have $g \p_g H(gA) > H(gA)$, and
moreover, by Appendix \ref{app_measconc}, the measure is concentrated on the Gribov horizon
where $H(gA) = (N^2-1)dV$.  This gives, in the limit $V \to \infty$,
\beqa
    \mV(0) & > &   \int_{|\vk| < k_{\rm max}} { d^dk \over (2 \pi)^d
    }\left\langle  {  \vk^2  \over \lambda_{|\vk|}^2(gA) } \right\rangle >
    \int_{|\vk| < k_{\rm max}} { d^dk \over (2 \pi)^d } \vk^2 \left\langle  {
    1 \over \lambda_{|\vk|}(gA) } \right\rangle^2 
\eeqa
Suppose now that $\langle  1 / \lambda_{|\vk|}(gA) \rangle$ has a power-law
behavior:
\beq
    \langle  1 / \lambda_{|\vk|}(gA) \rangle \approx b/ | \vk |^{2 +
    \eta},
\eeq
where $\eta > 0$ because of the horizon condition.  (We shall show in
the next section that this is really an ansatz for the infrared behavior of the
ghost propagator.)  It follows that
\beq
    \mV(0) >  \int_{|\vk| < k_{\rm max}} { d^dk \over (2 \pi)^d } \ { b^2 \over
    |\vk|^{2 + 2\eta} }.
\eeq
The color-Coulomb potential is confining if the self-energy $\mV(0)$ is
infrared divergent.  For the physical case of $d = 3$ dimensions, a sufficient
condition for this is $\eta \geq 1/2$.  

The origin of a long-range color-Coulomb force is now clear.  The horizon
condition assures that $\eta$ is positive and, as we shall see shortly, this
corresponds to a long range of the ghost propagator $G(r)$.  The last
inequality implies that $\mV(0)$ is more singular than $\mG(0)$ and this
assures that the color-Coulomb potential $\mV(r)$ is longer range than the
ghost propagator $G(r)$.

So far we have only used properties of the eigenvalues $\lambda_n(gA)$ of the
Faddeev-Popov operator $\mM$.  We shall obtain more detailed information about
the color-Coulomb potential $\mV(r)$ by a more detailed investigation of the
ghost propagator $\mM^{-1}(gA)$ in the presence of a background gluon field
$A$.

\section{The Ghost Propagator}%

Consider the spectral decomposition of the ghost propagator in a background
gauge field,
\beqa
\label{diagFT}
        \widetilde \mG(\vp; gA) & \equiv & {1 \over V } \int d^dx d^dy
        \ \exp[-i \vp \cdot (\vx - \vy) ] \ \mM^{-1}(\vx, \vy; gA)
         \nonumber  \\ & = & \sum_{|\vk| }
                { \langle  \vp |  P_{ | \vk | }(gA) | \vp \rangle
                \over \lambda_{|\vk |}(gA)}.
\eeqa
The projector has been calculated in Appendix \ref{app_projector}, eq.~\eqref{finalPa}, with the
result
\beq
\label{N+f}
    \langle \vp | P_{ | \vk | }(gA) | \vp \rangle = \langle \vp | W^2_{ | \vk |
    }(gA) | \vp \rangle \ \delta_{ \vp; | \vk | } + V^{-1}\int d^dz \ \left|
    \widetilde Z_{|\vk| \vp \vec z}(gA) \right|^2 (1 -  \delta_{ \vp; | \vk | } ), 
\eeq
where 
\beq
    \widetilde Z_{|\vk| \vp \vec z}(gA) \equiv \int d^dx \ \exp(-i \vp \cdot
    \vx) \ Z_{|\vk| \vx \vec z}(gA),
\eeq
$W_{|\vk| \vx \vec z}(gA)$ and $Z_{|\vk| \vx \vec z}(gA)$ are defined in
\eqref{defineW} and \eqref{defineZ}, and $\delta_{ \vp; | \vk | } = 1$ if
$\vp^2 = \vk^2$ and $\delta_{ \vp; | \vk | } = 0$ otherwise.  The cross terms
vanish, $\langle \vp | Z_{ | \vk | }(gA) W_{ |\vk | }(gA) | \vp \rangle =
\langle \vp | W_{ | \vk | }(gA) Z_{ | \vk | }^\dag(gA) | \vp \rangle = 0$,
because either $| \vp \rangle$ lies in the subspace $ P_{ | \vk | }(0)$ or it
lies in the orthogonal subspace $ Q_{ | \vk | }(0)$, but not both.  The first
term in \eqref{N+f} represents the ``probability" that the exact state $P_{ |
\vp | }(gA)$ remains in the unperturbed multiplet $P_{ | \vk | }(0)$, and has
the value
\beq
    \langle \vp | W^2_{ | \vk | }(0) | \vp \rangle = 1
\eeq
at $g = 0$, while the second term represents the probability of transition from
the unperturbed multiplet $P_{ | \vk | }(0)$ to a {\em different} unperturbed
multiplet $P_{ | \vp | }(0)$, with $|\vp| \neq |\vk|$.  Both terms in \eqref{N+f} are regular
because of the sum-rule \eqref{sumto1}.  We have $Z_{ |
\vk | }(gA) = O(\vk)$, so the second term is of order $\vk^2$, and we write
\beq
    \vk^2 B_{|\vk| \vp}(gA) \equiv \int d^dz \ \left| \widetilde Z_{|\vk| \vp
    \vec z}(gA) \right|^2,
\eeq
where $B_{|\vk| \vp}(gA)$ is regular. This follows from the fact that the sum
rule \eqref{sumto1} implies that $V^{-1} \Sigma_p B_{|\vk| \vp}(gA) \leq 1$.
Since there are of order $V$ terms that contribute significantly to the sum,
each being positive definite, $B_{|\vk| \vp}(gA)$ is expected to be order 1.
This gives
\beq
    \widetilde{\mG}(\vp; gA) = { \langle \vp | W^2_{ | \vp | }(gA) | \vp
    \rangle \over \lambda_{|\vp|}(gA) } + {1 \over V} \sum_{|\vk| \neq |\vp|}
    {\vk^2 B_{|\vk| \vp}(gA) \over \lambda_{|\vk|}(gA) }.
\eeq

We now take the infinite volume limit
\beq
    \widetilde{\mG}(\vp; gA) = { \langle \vp | W^2_{ | \vp | }(gA) | \vp
    \rangle \over \lambda_{|\vp|}(gA) } + \int {d^dk \over (2 \pi)^d} {\vk^2
    B_{|\vk| \vp}(gA) \over \lambda_{|\vk|}(gA) }.
\eeq
and look for the leading term for $\vp$ small.  We have $W^2_{ | \vp | }(gA) =
P_{ | \vp | }(0) + O(\vp^2)$ by \eqref{Worderk2}. The integral in the second
term converges in the infrared as long as $\lambda_{|\vk |}(gA)$ does not
vanish as rapidly as $|\vk|^{2 + d}$.  This gives for the leading term at small
$\vp$,
\beq
    \label{G_result}
    \widetilde{\mathcal{G}}^{ab}(\vp; gA) \approx {\langle \vp | P_{ | \vp |
    }^{ab}(0) | \vp \rangle \over \lambda_{|\vp|}(gA) } = { \delta^{ab} \over
    \lambda_{|\vp|}(gA) }.
\eeq
Here we have restored the color indices that have been previously suppressed
for simplicity. Note the color structure of $\widetilde \mG^{ab}$ is simply
$\delta^{ab}$ because the spectral decomposition of the long wavelength modes
has eigenvectors that don't feel the background gauge field, whereas the
eigenvalues are strongly perturbed, thus the color structure of $A^a$ only shows
up at higher orders in $k$. This makes manifest the equivalence of the
`no-pole' condition ($\widetilde \mG \to \infty$) and the `horizon condition'
($\lambda \to 0$) for $\vp \to 0$. 

\section{A Relation Between the color-Coloumb Potential and Ghost Propagator}%

 From \eqref{relateoperators} we get
\beq
    \widetilde \mV(\vk; gA) = \left( g \p_g + 1 \right)
    \widetilde \mG(\vk; gA) \approx  \left( g \p_g \right) {1 \over
    \lambda_{|\vk |}(gA) } = { - 1 \over \lambda_{|\vk |}^2(gA) }  g
    \p_g \lambda_{|\vk |}(gA).
\eeq  
where $\approx$ means the equality holds for the leading infrared behavior. 
We now establish some simple bounds that show intuitively how a long range
color-Coulomb potential arises.  To leading order in $\vk$ from
\eqref{evalue_eqn} we have
\beq 
        \widetilde \mV(\vk; gA)  \approx  { 1 \over
        (N^2 -1)d V }  { g \p_g H(gA) \  \vk^2  \over
        \lambda_{|\vk|}^2(gA) }.
\eeq 
According to \eqref{ehorizonfunction} we have
\beq
g \p_g H(gA) > H(gA),
\eeq
and moreover, by Appendix \ref{app_measconc}, the measure is concentrated on
the Gribov horizon where $H(gA) = (N^2-1)dV$.  This gives, 
\beq
    \widetilde \mV(\vk; gA) >  { \vk^2  \over \lambda_{|\vk|}^2(gA) },
\eeq
and for the Fourier-transform of the color-Coulomb potential
$\widetilde\mV(\vk) = \langle \widetilde \mV(\vk; gA) \rangle$ we
have
\beqa
\label{inequality}
    \widetilde \mV(\vk) & > &   \left\langle  {\vk^2  \over
    \lambda_{|\vk|}^2(gA) } \right\rangle > \ \vk^2 \left\langle  {  1
    \over \lambda_{|\vk|}(gA) } \right\rangle^2 \nonumber \\ & \approx &
    \vk^2 \left\langle  \widetilde \mG(\vk; gA) \right\rangle^2 = \
    \vk^2 \mG^2(\vk),
\eeqa
where we have used \eqref{G_result}.\footnote{This strikes of a similar inequality that
holds for all $k$, shown to the authors by Wolfgang Schleifenbaum in a private
communication.  Consider the following positive definite expression
\beqa
\langle (\widetilde{\mG} (\vec{k}; gA) -
\mG(\vec{k}))(\vec{k}^2)(\widetilde{\mG} (\vec{k}; gA) - \mG(\vec{k}))
\rangle & = \nonumber \\ \langle (i\vec{k} (\widetilde{\mG}(\vec{k}; gA) -
\mG(\vec{k}))^{\dag}(i\vec{k}(\widetilde{\mG}(\vec{k}; gA) - \mG(\vec{k})))
    \rangle & \geq 0  \; . \nonumber
\eeqa
By expanding this expression, one obtains:
\beq
\langle \widetilde{\mG}(\vec{k};gA) (\vec{k}^2) \widetilde{\mG}(\vec{k}; gA)\rangle \geq \vec{k}^2 \widetilde{\mG}^2(\vec{k}) \; . \nonumber
\eeq
This is not quite the same as \eqref{inequality} because
$\widetilde{\mV}(\vec{k})$, defined as the diagonal Fourier transform of the
color-Coulomb potential, is not the product of the diagonal Fourier transform of
each operator that makes up $\widetilde{\mV}(\vec{k})$. Schematically,
\beq
\widetilde{\mV} = \langle \mathcal{F}\left[ M^{-1} (-\p^2) M^{-1} \right]\rangle \neq
\langle
\mathcal{F}\left[M^{-1}\right]\mathcal{F}\left[-\p^2\right]\mathcal{F}\left[M^{-1}\right]
\rangle \; , \nonumber
\eeq
where $\mathcal{F}\left[f\right]$ is the diagonal Fourier transform \eqref{diagFT}.
Our inequality holds only for small $\vec{k}$ suggesting that this operator
does infact become simply a product in the IR.  This follows from the fact that
in the spectral decomposition of $\widetilde{\mM}^{-1}$, the projection
operator onto the eigenstates becomes diagonal in Fourier space for small
$\vec{k}$.}

Suppose now that the ghost propagator
$\mG(\vk)$ has a power-law behavior $\mG(\vk) \approx b/
|\vk|^{2 + \eta}$.  It follows that
\beq
    \widetilde\mV(\vk) >  { b^2 \over |\vk|^{2 + 2\eta} }.
\eeq

This gives for the color-Coulomb potential at large $r$
\beq
    \mV(r) > r^{2 + 2\eta - d}.
\eeq
For the physical case, $d = 3$, the term on the right is confining if $\eta >
1/2$, consistent with our previous result, and linearly rising if $\eta = 1$.

\section{Gauge Invariant Aspect of the Confining Force}%

The operator $\mM(gA) = - \p_i D_i(gA)$ is not gauge invariant,
nor is the color-Coulomb potential $\mV(r)$ so, while it is perfectly
legitimate to choose a gauge that is convenient for calculations, it is
instructive that we may give a gauge-invariant characterization of the
near-zero modes.

Consider the instantaneous correlator $K(\vx, \vy; gA)$ (see Appendix
\ref{app_kugo}) in a background field 
\beq
    \left\langle (D_i c)^a(x) (D_i \bar c)^b(y) \right\rangle_A = \delta^{ab}
    K(\vx, \vy; gA) \delta(x_0 - y_0), 
\eeq
where $D_i^{ac} \equiv \delta^{ac} \p_i + gf^{abc} A_i^b$ is the
gauge-covariant derivative, and
\beq
K(\vx, \vy; gA) \equiv D_{i \vx}^{ab} D_{i \vy}^{ac}
    (\mM^{-1})^{bc}(\vx, \vy; gA).
\eeq
Its diagonal Fourier transform is given by
\beq
    \widetilde K(\vp; gA) \equiv {1 \over V } \int d^dx d^dy \
    \exp[i \vp \cdot (\vx - \vy) ] \ K(\vx, \vy;
    gA),
\eeq
which has the spectral decomposition  
\beq
\label{DcDcbar}
    \widetilde K(\vp; gA) = \sum_{\vk} {  \langle  \vp |
    D_i\psi_{\vk} \rangle \langle  D_i\psi_{\vk} |  \vp \rangle
    \over \lambda_{ | \vk | }(gA)},
\eeq
where
\beq
    \langle \vp | D_i \psi_{\vk} \rangle \equiv V^{-1/2} \int d^dx \
    \exp(i \vp \cdot \vx) \ D_i \psi_{\vk}(\vx).
\eeq
This quantity is finite in the infrared limit where it is given by
\beq
\label{boundFof0}
    \widetilde K(0; gA) = {H(gA) \over V } \leq (N^2 - 1) d.
\eeq
which follows from \eqref{horizonfna}.

We return to \eqref{DcDcbar} and sum over $\vp$,
\beq
    {1 \over V} \sum_{\vp} \widetilde K(\vp; gA) = {1 \over V } \sum_{\vk} {
    || D_i\psi_{\vk} ||^2  \over \lambda_{ | \vk | }(gA)},
\eeq
where $  || D_i\psi ||^2 = \int d^dx \ | D_i\psi^a(\vx)|^2$ is the square
norm of $D_i \psi$.  The sum on $\vp$ is finite in the infrared, as we have
just demonstrated, and it can be made finite in the ultraviolet by introducing
a lattice cut-off \cite{Zwanziger:1991ac}.   
Because $\lambda_{|\vk|}(gA)$ vanishes more rapidly then $k^2$ when $A$ lies on
the Gribov horizon, this puts strong constraints on the form of $||D_i \psi_{\vk}||$
at small $k$. For example, suppose the configuration $A$ lies on the Gribov
horizon, where the measure is concentrated. We have $\lambda_{\vk}(gA) =
c(gA) |\vk|^{2 + \xi(gA)}$, where $\xi(gA)$ is the leading $k$ dependence of
$j(gA)$ from \eqref{evalue_eqn}. This gives, in the infinite-volume limit, 
\beq
    \int d^dp \ \widetilde K(\vp; gA) = \int d^dk \ { ||
    D_i\psi_{\vk}(gA) ||^2  \over c(gA) |\vk|^{2 + \xi(gA)} }
    \hspace{2cm}  {\rm for} \ A \in \p \Omega .
\eeq
From \eqref{boundFof0} the integral on the left-hand side is finite. We
conclude that if $\xi(gA) \geq d - 2$, then\footnote{It should be possible, at
least in principle, to verify this equation by numerical simulation on
sufficiently large lattices.} 
\beq
\label{gaugeinvarianteq}
    \lim_{\vk \to 0} || D_i \psi_{\vk}(gA) || = 0,
\eeq
while $ || \psi_{\vk} || = 1$. For $d = 2$ this holds if $\xi(gA) \geq 0$,
which is true by definition. For $d = 3$ this holds if $\xi(gA) \geq 1$. 

Thus in the infinite-volume limit, if $\xi(gA) \geq d -2$ holds, then {\em all}
configurations $A$ on the Gribov horizon are such that the gauge-covariant
derivative $D_i(gA)$ possesses a non-trivial zero-mode $D_i \psi(gA) = 0$. This
inequality is not unreasonable if one expects the power law behavior of
$\lambda_{|\vk|}(gA)$ on the horizon to be similar to that of its expection
value $\xi(gA) \sim \eta$.  For $d = 3$, $\eta \geq 1$ corresponds to at least
a linearly rising CCP, which is found for its lower bound, i.~e.\ the Wilson
potential, on the lattice\footnote{From numerical studies
        \cite{Greensite:2004ke}, it appears that the color-Coulomb potential is
        confining in the high-temperature, deconfined phase, and
\eqref{gaugeinvarianteq} holds more generally.}.

On the other hand, {\em no} configuration $A$ in the interior of the Gribov
region satisfies this condition because it implies $\p_i D_i \psi(gA) = 0$,
which is the condition for $A$ to be on the Gribov horizon.  We conclude that
in the infinite-volume limit, the measure is concentrated on configurations $A$
that support the zero mode $D_i \psi(gA) = 0$.  Thus by fixing a gauge, we have
nevertheless arrived at a gauge-invariant conclusion.

These gauge orbits are degenerate in that they lack at least one dimension.
Indeed the infinitesimal gauge transformation $g = 1 + \epsilon \psi$ generated
by $\psi$ leaves $A$ invariant, $A_i + \epsilon D_i(A) \psi =  A_i$. Also,
it can be easily shown that connections that support a non-trivial zero mode
are gauge transformations of abelian configurations which unifies the
concentration of measure hypothesis with the notion of `abelian dominance'
\cite{PhysRevD.25.2681}.  

\section{Comparison with the Lattice Data}%

Reference \cite{Nakagawa:2010eh} provides a detailed diagnostic of the
eigenfunction expansion of the color-Coulomb potential. It contains a number of
observations that accord qualitatively with the results obtained here. Namely,
the color-Coulomb potential $\mV(r)$ at large $r$ is well approximated by (i) a
comparatively small number of the infrared modes, (ii) among these it is well
approximated by the diagonal terms in the expansion \eqref{modeexpV}, and (iii)
all horizons are one horizon meaning that the low-lying modes
$\lambda_{|\vk|}(gA)$ pass through zero together (see Fig.~\ref{one_horizon}).

Indeed from \eqref{defineNinverse} we obtain the alternative expansion of the
color-Coulomb potential in the external field
\beq
    \left( \mN^{-1} \right)_{xy}^{ab} = \int d^dz \sum_m { \psi_{mx}^a
    \psi_{mz}^c \over \lambda_m } \left(-\p_z^2\right) \sum_n {
    \psi_{nz}^c \psi_{ny}^b \over \lambda_n } = \sum_{n,m} \omega_{mn}
    \psi_{mx}^a \psi_{ny}^b ,
\eeq
where
\beq
    \omega_{mn} = {\int d^dz \ \psi_{mz}^c (- \p_z^2) \psi_{nz}^c \over
    \lambda_m \lambda_n }.
\eeq
Upon comparison with \eqref{modeexpV} we have
\beq
    \omega_{nn} = { - g \p_g \lambda_n \over \lambda_n^2 } + { 1 \over\lambda_n
    } \hspace{2cm}  {\rm (no \ sum \ on \ } n)
\eeq
\beq
    \omega_{mn} =  { c_{nm} \over \lambda_n} + { c_{mn} \over \lambda_m}
    \hspace{2cm}  {\rm (for \ } m \neq n).
\eeq
It will be seen that the dominant term in the color-Coulomb potential comes
from $ - g \p_g \lambda_n / \lambda_n^2  $.  The eigenvalues $\lambda_n$ and
eigenfunctions $\psi_{nx}^a$ of the Faddeev-Popov operator $\mM$ and the
$\omega_{mn}$ have been evaluated numerically in \cite{Nakagawa:2010eh}.  

In Fig.~7 of \cite{Nakagawa:2010eh} it is observed that ``... the diagonal
components of the color-Coulomb potential are extremely large compared to the
off-diagonal components."  This is consistent in the infrared regime with our
result in Appendix \ref{app_projector} that the projector onto the eigenspace
--- that emerges from an unperturbed eigenvalue $\vk^2$ --- is the free
projector, $P_{|\vk|}(gA) = P_{|\vk|}(0)$. In the IR regime, the eigenfunctions
are approximately plane waves so the off-diagonal matrix elements of the
Laplacian operator vanish, $\omega_{\vk \vp} \sim \int d^d x \ \exp(- i \vk
\cdot \vx) (- \p^2) \exp( i \vp \cdot \vx) = 0 $, for $\vk \neq \vp$.  This
indicates that the diagonal elements alone should give a good approximation to
$\mV(r)$ at large separation $r$, in accordance with \eqref{modeexpV}. It would
be worthwhile investigating numerically whether the diagonal elements alone
provide a good approximation to $\mV$ at large $r$ or small $k$. We also
suggest that $|| D_i\psi_n(gA) ||$ be calculated to see if $\lim_{V \to
\infty}|| D_i\psi_n(gA) || \to 0$.  

Numerical data for low-lying eigenvalues of the Faddeev-Popov operator and for
long-range color-Coulomb potential $\mV(r)$  show sensitivity to choice of
Gribov copies \cite{Nakagawa:2010eh}.  The question then arises: which if any
numerical gauge choice corresponds to the GZ action with its horizon condition
\cite{Sternbeck:2015gia}
?  In numerical studies, gauge fixing is done by taking a
configuration $A$ generated by a standard gauge-invariant Monte Carlo
procedure, and minimizing the Hilbert square norm $|| A ||^2$ with respect to
local gauge transformations $g$, by some numerically convenient minimization
procedure so a relative minimum $A_1$ is achieved.  ``First copy" is whatever
local minimum $A_1$ the numerical algorithm finds.  To generate another copy, a
completely random gauge transformation  of $A_1$ is made, and the minimization
process is repeated.  This may be done a certain number of times, and what is
called the ``best" copy is the one that provides the lowest Hilbert norm.  

As described, the procedure generates configurations that satisfy the Landau
gauge condition and lie inside the Gribov region.  However if the minimization
is done separately on each time slice, and the minimum among gauge copies on
each time slice is chosen, then one has configurations that satisfy the Coulomb
gauge condition and lie inside the corresponding Gribov region\footnote{The
gauge that interpolates between Landau and Coloumb has been studied numerically
in \cite{Maas:2006fk}}. Among these
relative minima that are gauge copies of each other, one does {\em not} choose
the one with lowest Hilbert norm.  Instead, one calculates numerically the
lowest non-trivial eigenvalue $\lambda(A_i)$ of the Faddeev-Popov operator
$M(A_i)$ for each gauge copy $A_i$, and, to fix the gauge, one chooses that
configuration $A_n$ that provides the lowest non-trivial eigenvalue
$\lambda(A_n)$.  Thus one has $\lambda(A_n) \leq \lambda(A_i)$ for all gauge
copies $A_i$.

This procedure has been done in the case of Landau gauge in
\cite{Sternbeck:2012mf}, and something similar, where one chooses the copy that
gives the largest value of the ghost dressing function, $k^2 \mG(k)$
at $k=0$, was done in \cite{Maas:2009se}. Although in \cite{Sternbeck:2012mf}
there wasn't strong dependence of their gauge dependent quantities on which
Gribov copy was chosen, further studies on lattices with different volumes
should be done. Picking the Gribov copy closest to the horizon is likely to
be the best way to compare lattice data with calculations of gauge dependent
quantities using the GZ action due to the fact that the measure is concentrated
on the horizon as is shown in Appendix \ref{app_measconc}.

\section{Conclusion}%

The mechanism for a long range force in QCD in this approach goes as follows.
The ghost propagator $\mG(k)$ in Coulomb gauge is enhanced at $k = 0$ as compared
to $1/k^2$ by the horizon condition, which is identically satisfied in the GZ
formulation of QCD.  The relevance of the horizon condition to the singularity
of the ghost propagator $\mG(k) = \langle \widetilde{\mathcal{G}}^{ab}(\vk; gA)
\rangle$  is most clear in the exact relation \eqref{G_result}, obtained here,
$\widetilde{\mathcal{G}}^{ab}(\vk; gA)  = \delta^{ab} / \lambda_{|\vk|}(gA)$,
where $\lambda_{|\vk|}(gA)$ is the eigenvalue that emerges from
$\lambda_{|\vk|}(0) = k^2$.  From an analysis of the eigenvalues and
eigenspaces of the FP operator, we obtain an exact bound on the color-Coulomb
potential $\widetilde \mV(k) \geq k^2 \mG^2(k)$ which holds for $k \to 0$.  This
lower bound on $\widetilde \mV(k)$ implies that the color-Coulomb potential is
confining provided that the ghost propagator satisfies $\mG(k) \sim 1/ k^{2+
\eta}$ where $\eta \geq 1/2$.  It should be noted that $\mV(r)$ is a
gauge-dependent quantity, and the condition that it be confining is a necessary
condition for physical confinement but not sufficient.  This picture is quite
simple and provides an intuitive understanding of the mechanism of confinement.
Lastly, a gauge-invariant description of the Gribov horizon has been found,
which makes contact with other confinement scenarios.

Qualitative features of the present scenario have been observed in the lattice
simulation of \cite{Nakagawa:2010eh} which has been very encouraging for the
work presented here.  We believe that quantitive deviations could be due to the fact
that the numerical gauge fixing used in \cite{Nakagawa:2010eh} deviates from
the `analytic' gauge that corresponds to the GZ action with its horizon
condition.    

We are currently engaged in finding a solution to the Schwinger-Dyson equations
which promises to satisfy both the bound obtained here and the bound which
assures that there is no confinement without Coulomb confinement.

\begin{center}
    \line(1,0){125}
\end{center}

ACKNOWLEDGMENTS.  We recall with pleasure many stimulating conversations with Martin Schaden, Wolfgang Schleifenbaum, and Geoff Ryan.

\appendix
\section{Eigenvalues of the Faddeev-Popov Operator}%
\label{evalue_app}                                             %

We wish to find the eigenvalues of the Faddeev-Popov operator $\mM$ and
to evaluate the projector
\beq
\label{projectorP(gA)}
    P_{| \vk |}(gA) \equiv \sum_{ \ \vk; |\vk |} | \psi_{\vk}(gA) \rangle
    \langle \psi_{\vk}(gA) |
\eeq
that projects onto the space of eigenvectors $\psi_{\vk}(gA)$ of
$\mM$ that evolves from the degenerate unperturbed state with projector
\beq
    P_{| \vk |}(0) \equiv \sum_{ \ \vk; |\vk |} | \vk \rangle
    \langle \vk |.
\eeq
Here $\sum_{ \ \vk; |\vk |}$ means sum on all vectors $\vk = (2 \pi
\vec n/ L)$, with the same fixed $| \vk |$, where the $n_i$ are integers, i = 1, ... d, and the quantization volume $V = L^d$.  The states have a color index which takes $N^2 - 1$ values that is suppressed
for simplicity.  We call $V_{| \vk |}(0)$ the linear vector space, of
dimension $T = (N^2 - 1) \sum_{\vk; | \vk |} 1$, onto which $P_{|
\vk |}(0)$ projects. We follow the method of \cite{Picasso:2009}, however we only use
exact equations and not the perturbative expansion.

As a first step we seek an operator $S_{| \vk |}(gA)$ with the property
\beq
\label{Sequation}
    \mM(gA) S_{| \vk |}(gA) = S_{| \vk |}(gA) \kappa_{| \vk
    |}(gA), 
\eeq
where $S_{| \vk |}(gA)$ and $\kappa_{| \vk |}(gA)$ satisfy
\beq
    S_{| \vk |}(gA) P_{| \vk |}(0) = S_{| \vk |}(gA)
\eeq
\beq
    P_{| \vk |}(0) \kappa_{| \vk |}(gA) P_{| \vk |}(0) = \kappa_{|
    \vk |}(gA).
\eeq
We also require
\beq
\label{startingpoint}
    S_{| \vk |}(0) = P_{| \vk |}(0),    \hspace{3cm}  \kappa_{| \vk
    |}(0) = P_{| \vk |}(0) \vk^2.
\eeq
The operator $S_{| \vk |}(gA)$ effects a block diagonaliztion of the
Faddeev-Popov operator $\mM(gA)$ onto a block $\kappa$ of dimension~$T
\times T$, and $\kappa$ is an operator that acts within $V_{| \vk |}(0)$.
As $g$ increases from 0 to some finite value, the eigenspace of $P_{| \vk
|}(gA)$ evolves from the degenerate level with eigenvalue $\vk^2$ as shown in
Fig.~\ref{perturbed_evalue}.  

We try a solution $S_{| \vk |}(gA)$ of the form
\beq
    S_{| \vk |}(gA) = P_{| \vk |}(0) + R_{| \vk |}(gA),
\eeq
where $R_{| \vk |}(gA)$ maps the unperturbed subspace into the orthogonal
subspace with projector $Q_{| \vk |}(0) = I - P_{| \vk |}(0)$, so
\beq
\label{PzeroonR}
    Q_{| \vk |}(0) R_{| \vk |}(gA) P_{| \vk |}(0) = R_{| \vk
    |}(gA),
\eeq
and moreover
\beq
\label{PR=0}
    Q_{| \vk |}(0) P_{| \vk |}(0) = P_{| \vk |}(0) Q_{| \vk
    |}(0)  = 0.
\eeq

Equation \eqref{Sequation} reads 
\beq
\label{Requation}
    [\mM_0 + \mM_1(gA)] [P_{| \vk |}(0) + R_{| \vk
    |}(gA)] = [ P_{| \vk |}(0) + R_{| \vk |}(gA)] \kappa_{| \vk
    |}(gA). 
\eeq
We apply the orthogonal projectors $P_{| \vk |}(0)$ and $Q_{| \vk
|}(0)$ separately to \eqref{Requation}, and obtain
\beqa
\label{eqsforkap}
    \vk^2 P_{| \vk |}(0) + P_{| \vk |}(0) \mM_1 P_{|
    \vk |}(0) + P_{| \vk |}(0) \mM_1 R_{| \vk |}(gA) & = &
    \kappa \\   \label{eqsforR} Q_{| \vk |}(0) \mM R_{| \vk |}(gA) + Q_{| \vk
    |}(0) \mM_1 P_{| \vk |}(0) & = & R_{| \vk |}(gA) \kappa,
\eeqa
where we have used $P_{| \vk |}(0) + Q_{| \vk |}(0) = I$, and $P_{| \vk
|}(0)Q_{| \vk |}(0) = Q_{| \vk |}(0) P_{| \vk |}(0) = 0$.  The matrix $\kappa$
could be diagonalized by a unitary transformation $\kappa = u \lambda_{\rm d}
u^\dag$, where $\lambda_{\rm d}$ is a $T$-dimensional diagonal matrix and $u$
is a $T$-dimensional unitary matrix.  We now let the volume $V = L^d$ approach infinity,
and we shall assume that the spectrum becomes continuous in this limit for
configurations $A$ that lie inside the Gribov horizon. We shall also assume
that there is no level crossing in the sense that if $|\vk| < |\vk'|$ then
$\lambda_{|\vk|} < \lambda_{|\vk'|}$. In this case $\lambda_d$ approaches
$\lambda_{| \vk |}(gA) P_{| \vk |}(0)$, where $\lambda_{| \vk |}$ is a number,
$\kappa = \lambda_{| \vk |}(gA) u P_{| \vk |}(0) u^\dag = \lambda_{| \vk |}(gA)
P_{| \vk |}(0)$, and \eqref{Sequation} reads
\beq
\label{Sequationc}
    \mM(gA) S_{| \vk |}(gA) = \lambda_{| \vk |}(gA) S_{|
    \vk |}(gA). 
\eeq

We formally solve \eqref{eqsforR} for $R_{| \vk |}(gA) = Q_{| \vk |}(0)
R_{| \vk |}(gA) P_{| \vk |}(0)$,
\beq
\label{Rsolution}
    R_{| \vk |}(gA) = - \left(\mM_{Q} - \lambda_{| \vk |}(gA)
    \right)^{-1} Q_{| \vk |}(0) \mM_1 P_{| \vk |}(0),
\eeq
where $ \mM_{Q} \equiv Q_{| \vk |}(0) \mM Q_{| \vk
|}(0)$. When substituted into \eqref{eqsforkap} for $\kappa$, this gives the
$T$-dimensional matrix equation
\beq
    \lambda_{| \vk |}(g)P_{| \vk |}(0) = \vk^2 P_{| \vk |}(0) +
    P_{| \vk |}(0) \mM_1 P_{| \vk |}(0) - P_{| \vk |}(0)
    \mM_1 \left(\mM_{Q} - \lambda_{| \vk |}(gA)
    \right)^{-1} Q_{| \vk |}(0) \mM_1 P_{| \vk |}(0).
\eeq
Upon taking the trace of this matrix, we obtain
\beq
    \lambda_{| \vk |}(gA) = \vk^2 + T^{-1} \sum_{\vk; |\vk|}
    \langle \vk | \mM_1^{aa} | \vk \rangle - L_{| \vk
    |}(gA),
\eeq
where
\beq
    L_{| \vk |}(gA) \equiv (VT)^{-1} \sum_{\vk; |\vk|} \int d^dx
    d^dy \ \exp(- i \vk \cdot \vx ) \left[\mM_1
    \left(\mM_Q - \lambda_{| \vk |}(gA) \right)^{-1}Q_{| \vk
    |}(0)  \mM_1\right]_{\vx, \vy}^{aa} \exp( i \vk \cdot
    \vy).
\eeq
We have $\mM_1^{ac} = -  f^{abc} g A^b \p_i = - \mM_1^{ca}$, so
the trace on color indices vanishes, $\mM_1^{aa} = 0$, and we obtain
\beq
    \lambda_{| \vk |}(gA) = \vk^2 - L_{| \vk |}(gA).
\eeq
The ghost momentum factors out when the operator $\mM_1^{ac}$ acts on
the plane wave $\exp(i \vk \cdot \vy)$
\beq
\label{factorghostmom}
    \mM_1^{ac} \exp(i \vk \cdot \vy) = - f^{abc} gA_i^b
    \p_i \ \exp(i \vk \cdot \vy) =  - i k_i f^{abc} gA_i^b \
    \exp(i \vk \cdot \vy),
\eeq
and correspondingly for $\vx \to \vy$, which gives
\beq
\label{Lsubka}
    L_{| \vk |}(gA) = T^{-1} \sum_{\vk; |\vk|} k_i k_j
    Y_{ij}^{aa}(k; gA),
\eeq
where
\beq
\label{Y1}
    Y_{ij}^{ab}(k; gA) \equiv V^{-1} \int d^dx d^dy \ \exp[ i \vk \cdot(
    \vy - \vx ) ] f^{abc} gA_i^b(x) f^{ade} gA_j^d(y) \left[
    \left(\mM_{Q} - \lambda_{| \vk |}(gA) \right)^{-1}  Q_{|
    \vk |}(0) \right]_{\vx, \vy}^{ce}.
\eeq
The operator $f^{ade} gA_j^d(y)$ appears in the sandwich
\beq
    Q_{| \vk |}(0)  f^{ade} gA_j^d(y) \exp(i \vk \cdot \vy).
\eeq
This allows us to make the substitution
\beq
\label{substitution}
    f^{ade} gA_j^d(y) \to \p_j^{(y)} \delta^{ae} + f^{ade} gA_j^b(y) =
    D_j^{(y)ae}
\eeq
(and correspondingly for $\vy \to \vx$), for we have
\beq
    Q_{| \vk |}(0) \ \p_j^{(y)} \ \exp(i \vk \cdot \vy) = ik_y \
    Q_{| \vk |}(0) \ \exp(i \vk \cdot \vy) = 0,
\eeq
which vanishes because $Q_{| \vk |}(0)$ is the projector onto the space
orthogonal to $\exp(i \vk \cdot \vy)$.  This gives
\beq
\label{Y2}
    Y_{ij}^{ab}(k; gA) = V^{-1} \int d^dx d^dy \ \exp[ i \vk \cdot( \vy
    - \vx ) ] D_i^{(x)ac} D_j^{(y)be} \left[  \left(\mM_{Q} -
    \lambda_{| \vk |}(gA) \right)^{-1}  Q_{| \vk |}(0)
    \right]_{\vx, \vy}^{ce}.
\eeq
The eigenvalue $\lambda_{| \vk |}(gA)$ is the solution of the equation
\beq
\label{factorksquarea}
    \lambda_{| \vk |}(gA) = \vk^2 \left( 1 -  T^{-1} \sum_{\vk;
    |\vk|} \hat k_i \hat k_j \ Y_{ij}^{aa}(\vk; gA) \right),
\eeq
where $Y_{ij}^{ab}(k; gA)$ depends on $\lambda_{| \vk |}(gA)$.
Configrations $A$ that lie on the Gribov horizon may be found by setting
$\lambda_{|\vk|}(gA) = 0$ in this equation.

\section{Projector onto States Emerging From Degenerate Subspace}%
\label{app_projector}                                                        %

We wish to represent the projector \eqref{projectorP(gA)} by
\beq
        P_{| \vk |}(gA) = U_{| \vk |}(gA)  U_{| \vk |}^\dag(gA),
\eeq
where $U_{| \vk |}(gA)$ satisfies
\beq
        U_{| \vk |}(gA) P_{| \vk |}(0) = P_{| \vk |}(gA) U_{|
        \vk |}(gA) = U_{| \vk |}(gA),
\eeq
and thus maps the $T$-dimensional vector space $V_{| \vk |}(0)$, which is
the eigenspace of $P_{| \vk |}(0)$, onto the $T$-dimensional vector space
$V_{| \vk |}(gA)$, which is the eigenspace of $P_{| \vk |}(gA)$.  Here
$T = (N^2 -1) \sum_{ \ \vk; |\vk |} 1 $ is the degeneracy of the
unperturbed level $\lambda_{|\vk|}(0) = \vk^2$.

Properties \eqref{Sequation} through \eqref{startingpoint} do not fix
$S_{| \vk |}(gA)$ uniquely, and we may multiply $S_{| \vk |}(gA)$ on the
right by any operator $W_{| \vk |}(gA)$ that acts within $V_{| \vk
|}(0)$,
\beq
\label{Wproperty}
    P_{| \vk |}(0) W_{| \vk |}(gA) P_{| \vk |}(0) = W_{| \vk
    |}(gA),
\eeq
and we may substitute
\beq
    S_{| \vk |}(gA) \to U_{| \vk |}(gA) = S_{| \vk |}(gA) W_{|
    \vk |}(gA).
\eeq
This freedom is a generalization of the fact that an eigenvalue equation does
not fix the normalization of the eigenfunction.  We now seek a hermitian, $W_{|
\vk |}(gA) = W_{| \vk |}^\dag(gA) $, so chosen that the operator
\beq
\label{definePi}
    P_{| \vk |}(gA) \equiv U_{| \vk |}(gA) U_{| \vk |}^\dag(gA) = S_{|
    \vk |}(gA) W^2_{| \vk |}(gA) S_{| \vk |}^\dag(gA),
\eeq
has the projector property
\beq
\label{projectorsat}
    P_{| \vk |}^2(gA) = P_{| \vk |}(gA).
\eeq
We have 
\beq
\label{Psquare}
    P_{| \vk |}^2(gA) = S_{| \vk |}(gA) W^2_{| \vk |}(gA) X_{| \vk
    |}(gA) W^2_{| \vk |}(gA) S_{| \vk |}^\dag(gA)
\eeq
where
\beq
\label{defineX}
    X_{| \vk |}(gA) \equiv S_{| \vk |}^\dag(gA) S_{| \vk |}(gA) = P_{|
    \vk |}(0) + R_{| \vk |}^\dag(gA) R_{| \vk |}(gA)
\eeq
and we have used \eqref{PzeroonR} and \eqref{PR=0}. 
It is clear
from \eqref{Psquare} that \eqref{projectorsat} is satisfied, as desired, by
choosing
\beq
    \label{defineW}
    W_{| \vk |}(gA) \equiv X_{| \vk |}^{-1/2}(gA).
\eeq
where the positive square root is understood.  Note that $X_{| \vk |}(gA)$
satisfies
\beq
\label{WinP(0)}
    P_{| \vk |}(0) X_{| \vk |}(gA)P_{| \vk |}(0) = X_{| \vk |}(gA)
\eeq
and is thus an operator that acts within $V_{| \vk | }(0)$,
so \eqref{Wproperty} is satisfied, as required, and we conclude that
\beqa
    U_{| \vk |}(gA) & = & S_{| \vk |}(gA) W_{| \vk |}(gA)
    \nonumber \\
    & = & W_{| \vk |}(gA) + R_{| \vk |}(gA) W_{| \vk |}(gA) .
\eeqa

We shall show that, in the low-momentum limit, the projector $P_{| \vk
|}(gA)$ remains unperturbed, $\lim_{\vk \to 0}P_{| \vk |}(gA) = P_{|
\vk |}(0)$.  Indeed, the expression for $R_{| \vk |}(gA)$, eq.
\eqref{Rsolution}, contains on the right the factor $\mM_1 P_{| \vk
|}(0)$, from which the ghost momentum factorizes as in \eqref{factorghostmom},
\beq
    [\mM_1 P_{| \vk |}(0)]_{\vx \vy}^{ac} = - f^{abc}
    gA_i^b \p_i  \ V^{-1} \sum_{\vk} \exp[i \vk \cdot ( \vx -
    \vy)] = - ik_i \ f^{abc} gA_i^b \ V^{-1} \sum_{\vk} \exp[i
    \vk \cdot ( \vx - \vy)],
\eeq
just as the external ghost momentum factorizes from Feynman diagrams in the
Coulomb and Landau gauges. It follows that
\beq
    R_{| \vk |}(gA) = O(\vk),
\eeq
and consequently also
\beqa
\label{Xorderk2}
    X_{| \vk |}(gA) = P_{|\vk |}(0) + O(\vk^2)
\nonumber \\
\label{Worderk2}
    W_{| \vk |}(gA) = P_{|\vk |}(0) + O(\vk^2)
\eeqa
For convenience we define
\beq
\label{defineZ}
 \  Z_{| \vk |}(gA) \equiv R_{| \vk |}(gA) W_{| \vk |}(gA),
\eeq
so
\beqa
 U_{| \vk|}(gA)  = W_{| \vk |}(gA) +  Z_{| \vk |}(gA).
 \nonumber 
\eeqa
It satisfies
\beqa
\label{Zprojectprop}
Q_{| \vk |}(0) Z_{| \vk |}(gA) P_{| \vk |}(0) & = & Z_{| \vk |}(gA)
\nonumber \\
Z_{| \vk |}(gA) & = & O(\vk).
\eeqa
We thus obtain from \eqref{definePi} for the projector
onto the space of states that emerges from $P_{| \vk |}(0)$
\beq
\label{finalPa}
    P_{| \vk |}(gA) = \left(W_{| \vk |}(gA) + Z_{| \vk
    |}(gA) \right) \ \left( W_{| \vk |}(gA) + Z_{| \vk |}^\dag(gA) \right).
\eeq

Note the identity
\beq
\int d^dx \ P_{| \vk | xx}^{ \ \  aa}(gA) = T
\eeq
where $T$ is the (finite) dimensionality of the degenerate level from which the
eigenstates emerged.  It implies
\beq
\label{sumto1}
    T^{-1}\int d^dx d^dy \ \left[ \left( W_{| \vk | xy}^{ \ \  ab}(gA)
    \right)^2 + \left( Z_{| \vk | xy}^{ \ \  ab}(gA) \right)^2 \right] = 1
\eeq
where  we have used $P_{| \vk |}(0) W_{| \vk |}(gA) P_{| \vk |}(0) = W_{| \vk
|}(gA)$ and $Q_{| \vk |}(0) Z_{| \vk |}(gA) P_{| \vk |}(0) = Z_{| \vk |}(gA)$.
Since both terms are positive, we may interpret each term respectively as the
``probability" that there is not, or there is, a transition out of  the
unperturbed vector space.  Neither term is divergent.  Moreover we have
\beq
\label{Porderk}
    P_{| \vk |}(gA) = P_{|\vk |}(0) + O(\vk),
\eeq
and we have
\beqa
\label{infraredlimitN}
    \lim_{\vk \to 0} X_{| \vk |}(gA) & = & \lim_{\vk \to 0} W_{| \vk |}(gA) =
    \lim_{\vk \to 0} P_{| \vk |}(gA) = P_{| \vk|}(0).
    \nonumber \\
    \lim_{\vk \to 0} R_{| \vk |}(gA) & = & \lim_{\vk \to 0} Z_{| \vk |}(gA) =
    O(\vk).
\eeqa
Thus there is no transition {\em out} of the unperturbed vector space in the
limit $\vk \to 0$, and we conclude that {\em in the infrared limit the
projector is unchanged} or, in terms of the eigenstates,
\beq
        \lim_{\vk \to 0} \sum_{ \ \vk; |\vk |} | \psi_{\vk}(gA)
        \rangle \langle \psi_{\vk}(gA) | = \sum_{ \ \vk; |\vk |} |
        \vk \rangle \langle \vk |.
\eeq
This is in stark contrast to the eigenvalues $\lambda_{|\vk|}(gA)$ which are
strongly perturbed as $A$ approaches the Gribov horizon.  In both cases the
leading correction vanishes with $\vk$.  However for the eigenvalue the leading
term is $\vk^2$, which vanishes with $\vk$, whereas for the projector, the
leading term is of order 1.

It may also be interesting to note that this result for the projector also
implies that the unknown function $F(\lambda)$ in 
\cite{Greensite:2004bv,Greensite:2004ur} in the infrared is simply given by $k^2$.

\section{The Horizon Condition}%
\label{app_horizon}

We now solve the exact equation \eqref{factorksquarea} for $\lambda_{| \vk
|}(gA)$ for small $\vk$.  We have
\beq
\label{factorksquareb}
    \lambda_{| \vk |}(gA) = \vk^2 \left( 1 -  T^{-1} \sum_{\vk;
    |\vk|} \hat k_i \hat k_j \ Y_{ij}^{aa}(0; gA) + j_{| \vk |}(gA)
    \right),
\eeq
where $ j_{| \vk |}(gA)$ is a remainder that vanishes with $\vk$,
$\lim_{\vk \to 0} j_{| \vk |}(gA) = 0$.  From
\beq
        T^{-1} \sum_{\vk; |\vk|} \hat k_i \hat k_j = [(N^2 -1) d]^{-1}
        \delta_{ij},
\eeq
we obtain
\beq
\label{factorksquarec}
    \lambda_{| \vk |}(gA) =  \vk^2 \left( 1 -  [(N^2 -1) dV]^{-1} \ H(gA) +
    j_{| \vk |}(gA) \right),
\eeq
where
\beq
H \equiv V  Y_{ii}^{aa}(0; gA)
\eeq
is the horizon function.  From \eqref{Y1} and \eqref{Y2}
we obtain two expressions for $H$,
\beq
\label{H1}
    H(gA) = \int d^dx d^dy \ f^{abc} gA_i^b(x) f^{ade} gA_i^d(y) \left(
    \mM_{Q}^{-1} \right)_{\vx, \vy}^{ce},
\eeq
and
\beqa
\label{H2}
    H(gA) = \int d^dx d^dy \ D_i^{(x)ac} D_i^{(y)be} \left( \mM_Q^{-1}
    \right)_{\vx, \vy}^{ce}.
\eeqa

Consider the projector $Q_{| \vk |}(0) = I - P_{| \vk |}(0)$ in a
position basis.  We have $\delta(\vx - \vy) = V^{-1} \sum_{\vk}
\exp[i \vk \cdot (\vy - \vx)] \to (2\pi)^{-d} \int d^d k \ \exp[i
\vk \cdot (\vy - \vx)]$.  On the other hand $[P_{| \vk
|}(0)]_{\vx, \vy} = V^{-1} \sum_{\vk; |\vk|} \exp[i
\vk \cdot (\vy - \vx)]$, which involves a finite sum, and we
have, in the infinite-volume limit, $\lim_{V \to \infty} P_{| \vk
|}(0) = 0$, and $\lim_{V \to \infty} Q_{| \vk |}(0) = I$.  This argument is correct when applied with sufficiently regular functions.  The second expression for $H(gA)$, which is closely related to the function introduced by Kugo and Ojima \cite{Kugo:1995km, Kugo01021979} that we shall consider shortly, is more regular than the first and gives
\beqa
\label{horizonfnz}
    H(gA) = \int d^dx d^dy \ D_i^{(x)ac} D_i^{(y)be} \left( \mM^{-1}
    \right)_{\vx, \vy}^{ce}.
\eeqa

Under the assumption that the condition of
no-level-crossing \cite{Neumann:1929} holds for configurations $A$ for which all
eigenvalues are positive, it follows that the first eigenvalue that goes
negative is the one that started out lowest at $gA = 0$, namely the one with
the lowest (non-zero) value of $\vk \to 0$.  This defines the (first) Gribov
horizon which, according to \eqref{factorksquarec} with $\lim_{\vk \to 0} j_{|
\vk |}(gA) = 0$, is given by
\beq
\label{horizonH}
    H(gA) = (N^2 -1) d V.
\eeq

\section{Relation to the Kugo-Ojima function}%
\label{app_kugo}

Coonsider the quantity
\beq
\label{Kij}
    K_{ij}^{ab}(\vk; gA) \equiv V^{-1} \int d^dx d^dy \ \exp[ i \vk \cdot(
    \vy - \vx ) ] D_i^{(x)ac} D_j^{(y)be} \left( \mM^{-1}
    \right)_{\vx, \vy}^{ce},
\eeq
which provides a regularization of the horizon function,
\beqa
\label{horizonfna}
    H(gA)/V & = & \lim_{\vk  \to 0} K_{ii}^{aa}(\vk; gA) \nonumber \\ &
    = & \lim_{\vk  \to 0} V^{-1} \int d^dx d^dy \ \exp[ i \vk \cdot(
    \vy - \vx ) ] D_i^{(x)ac} D_i^{(y)be} \left( \mM^{-1}
    \right)_{\vx, \vy}^{ce}.
\eeqa
Its expectation-value yields the function $u(\vk^2)$ introduced
in Landau gauge by Kugo and Ojima \cite{Kugo:1995km, Kugo01021979}
\beqa
    \langle K_{ij}^{ab}(\vk; gA) \rangle & = & V^{-1} \int d^dx d^dy \ \exp[ i
    \vk \cdot( \vy - \vx ) ] \langle D_i c(x) D_j \bar c(y)
    \rangle^{ab}, \nonumber \\ & = & \left[ - \left(\delta_{ij} - { k_i k_j \over
    \vk^2 } \right) u(\vk^2) + { k_i k_j \over \vk^2 } \right]
    \delta^{ab},
\eeqa
and we have
\beq
    \langle H(gA) \rangle = [ - (d -1)u(0) +1] (N^2-1) V.
\eeq
The Kugo-Ojima criterion for confinement $u(0) = -1$ is identical to the
horizon condition $\langle H(gA) \rangle = d (N^2-1) V$ and is thus
automatically satisfied in the GZ formulation of QCD.  

\section{Measure Sits on the Horizon}%
\label{app_measconc}

There remains to show that the measure of the horizon function $H(gA)$ is concentrated on
the Gribov horizon.

Let $g$ be a free positive parameter $g \geq 0$.  As $g$ varies for fixed $A$,
$gA$ defines a ray in $A$-space which intersects the Gribov horizon $\p
\Omega$ at a unique value defined by $H(gA) = (N^2-1) d  V$, which is the same
for each ray.  The Gribov region is convex, \cite{Zwanziger:1982na}, and we
have obviously $H(0) = 0$. 

{\em Lemma.}  In the interior of the Gribov region the horizon function $H(g
A)$ increases monotonically with $g$ along every ray $g A$ in $A$-space,
starting at $H(0) = 0$, and ending on the Gribov horizon defined by $H(gA) =
(N^2-1) d V$.  {\em Proof.} It suffices to show that the derivative $\p_g
H(gA)$ is positive.  From \eqref{H1} we have
\beqa
\label{dhorizonfunction}
    g \p_g H(g A) & =  \int d^d x \int d^d y & \Big[  2 f_{bac} gA_i^a (x) \
    (\mM_Q^{-1})^{c d}(x, y; g A) \ f_{bed} gA^e_i (y)
    \nonumber \\
    && \ \ \ \ - f_{bac} gA_i^a (x) \ \Big(  \mM_Q^{-1}( g A) \ ( - g
    A_j \p_j ) \ \mM_Q^{-1}(g A) \Big)^{c d}(x, y) \ f_{bed} gA^e_i (y)
    \Big] \;,
\eeqa
where $A_j$ acts in the adjoint representation.  We now use $ - g A_j
\p_j  = M(g A) + \p_j \p_j$, which gives
\beqa
\label{ehorizonfunction}
    g \p_g H(g A) & =  \int d^d x \int d^d y & \Big[ f_{bac} gA_i^a (x)
    \ (\mM_Q^{-1})^{c d}(x, y; g A) \ f_{bed} gA^e_i (y) \nonumber \\ && \ \
    \ \ +  f_{bac} gA_i^a (x) \ \Big(  \mM_Q^{-1}( g A) \ ( -  \p_j
    \p_j ) \ \mM_Q^{-1}(g A) \Big)^{c d}(x, y) \ f_{bed} gA^e_i (y)
    \Big] \;,
\eeqa
which is manifestly positive $g \p_g H(gA) > 0$ for $A \in \Omega$. QED 

In fact, as a {\em corollary}, which we used to derive the inequalities for the
color-coulomb potential, we also have, by \eqref{H1}, that $g \p_g H(gA) > H(gA)$.

{\em Theorem.}  In the limit of large volume $V$, the measure defined by the
non-local GZ action, 
\beq
    S_{GZ} = \int d^Dx \left[ \mathcal L_{\rm YangMills} + \mathcal L_{\rm FP} + \gamma
    (V^{-1} H(A) - d (N^2 - 1) ) \right]  \nonumber ,
\eeq
where the value of $\gamma$ is defined by the condition that the quantum effective
action be stationary with respect to it \cite{Vandersickel:2012tz}, is entirely supported on the boundary
$\p \Omega$ of the Gribov region.\\ {\em Proof.}  This is true because,
according to the horizon condition $\langle H \rangle = (N^2 - 1) d V =  {\rm
max}_{A \in \Omega} H(gA)$, the mean value of $H(A)$ equals the
maximum value that it achieves in the Gribov region, so it is supported where
the maximum is achieved.  This occurs on the boundary $\p \Omega$.  QED.  

Since the measure is concentrated on the Gribov horizon, the horizon
condition is in fact satisfied by almost all (in the probabilistic sense)
configurations that contribute to the expectation value, so even without
averages we make use of the identity $H(A) = d(N^2-1) V$.

\bibliography{bibliography}

\end{document}